\documentclass[10 pt]{IEEETran}

\def \be {\begin{equation}}
\def \ee {\end{equation}}
\def \nn {\nonumber}

\usepackage{ifpdf}

\usepackage[dvips]{graphicx}
\usepackage{color}

\usepackage[cmex10]{amsmath}
\usepackage{amssymb}

\title{\LARGE \bf Bilateral Control of Teleoperators with Closed Architecture and Time-Varying Delay}
\author{Hanlei~Wang, Yipeng Li, and Tiantian Jiang%\emph{IEEE Member}
\thanks{The authors are with the Science and Technology on Space Intelligent Control Laboratory,
Beijing Institute of Control Engineering,
Beijing 100094, China (e-mail: hlwang.bice@gmail.com; yipengli.bice@gmail.com; jiangtt@amss.ac.cn).}
}

\begin{document}

\maketitle
\thispagestyle{empty}
\pagestyle{empty}

\begin{abstract}
This paper investigates bilateral control of teleoperators with closed architecture and subjected to arbitrary bounded time-varying delay. A prominent challenge for bilateral control of such teleoperators lies in the closed architecture, especially in the context not involving interaction force/torque measurement. This yields the long-standing situation that most bilateral control rigorously developed in the literature is hard to be justified as applied to teleoperators with closed architecture. With a new class of dynamic feedback, we propose kinematic and adaptive dynamic controllers for teleoperators with closed architecture, and we show that the proposed kinematic and dynamic controllers are robust with respect to arbitrary bounded time-varying delay. In addition, by exploiting the input-output properties of an inverted form of the dynamics of robot manipulators with closed architecture, we remove the assumption of uniform exponential stability of a linear time-varying system due to the adaptation to the gains of the inner controller in demonstrating stability of the presented adaptive dynamic control. The application of the proposed approach is illustrated by the experimental results using a Phantom Omni and a UR10 robot.
\end{abstract}


% \fi
%\IEEEpeerreviewmaketitle

\section{Introduction}

In recent years, strong interest has been witnessed in the study of human-robot interaction, and one representative and also standard example is bilateral teleoperation with a teleoperator system. Typically, a teleoperator system involves two robots which are respectively referred to as the master and slave, and the master robot is usually guided or maneuvered by a human operator; if the interaction between the slave and the environment is reflected to the master as portion of its input torque, the teleoperator system is said to be controlled bilaterally \cite{Anderson1989_TAC}. Numerous control approaches have been proposed ever since the fundamental result based on the scattering theory in \cite{Anderson1989_TAC} (see, e.g., \cite{Lawrence1993_TRA,Hokayem2006_AUT}). The scattering-based approach is later systematically reformulated using wave variables in \cite{Niemeyer1991_JOE}. Other results addressing various aspects associated with bilateral teleoperation appear in, e.g., \cite{Niemeyer1998_ICRA,Yokokohji1999_IROS,Chopra2003_ACC,Hashtrudi-Zaad2002_TRA,Ryu2004_TRA,Stramigioli2005_TRO,Malysz2011_TRO}. Many synchronization-based approaches have also been developed in, e.g., \cite{Lee2006_TRO,Chopra2008_AUT,Nuno2009_IJRR,Wang2020b_AUT}, and in particular, delay-independent solutions are provided in \cite{Wang2018_CAC,Wang2020b_AUT} for bilateral teleoperation with unknown bounded time-varying delay. Most of these solutions, with rigorous design procedures and stability analysis, are formulated for teleoperators with an open torque design interface.

The line of industrial/commercial control systems, on the other hand, points towards a road that seems substantially different from the typical study of nonlinear control. For example, in most industrial/commercial robot manipulators, the torque design interface, the availability of which is typically required in the study of nonlinear control of robot manipulators, is not available. This, in some sense, yields the weak connection between control theory and industrial/commercial control applications, as is discussed in \cite{Wang2020_TAC}; rigorous nonlinear solutions have been presented in \cite{Wang2020_TAC} for regulation and tracking problems of industrial robots. Differing from the traditional automation, the current intensive research and interest in collaborative robots indicates a trend of modern automation with a variety of collaborative robots involving an environment that is often not so well structured. In this context, the robots generally need the guidance/intervening of human operators, which leads to semi-autonomous control systems and in particular semi-autonomous industrial/commercial control systems. Bilateral teleoperation is a representative mode for such systems (for instance, teaching an industrial robot by a human operator at a remote site).

In this paper, we focus on bilateral control of teleoperators involving robots with closed architecture (most industrial/commercial robots belong to this category and their torque design interface is typically not available) and subjected to arbitrary bounded time-varying delay and unavailability of force/torque measurement. The motivation, as is discussed above, is to enhance the impact of nonlinear bilateral control on industrial/commercial robots (and consequently modern automation), in contrast to most existing results concerning the experimental study of bilateral control approaches using robots with an open torque design interface (see, e.g., \cite{Rodriguez2009_TRO,Lee2010_TRO_PassiveSetPosition,Lee2006_TRO,Chopra2008_AUT}). Such a practice, if succeeded, would bridge the potential gap between the study of teleoperation control theory and practical robotic applications (beyond most current laboratory robotics research). The technical challenge of realizing this objective is mainly due to the closed architecture of the robots (i.e., no torque design interface is available and typically only the joint velocity or position command can be specified) and due to the typically employed proportional-derivative (PD) or proportional-integral-derivative (PID) control action of the inner control loop with unknown gains. This challenge also renders most existing results developed in the context of an open torque design interface (e.g., \cite{Anderson1989_TAC,Munir2002_TMECH,Lee2006_TRO,Chopra2008_AUT,Nuno2009_IJRR,Franken2011_TRO,Wang2020b_AUT,Nuno2018_IJRNC,Heck2018_TRO}) no longer applicable. The involved challenge for bilateral control of teleoperators subjected to closed architecture may be alleviated via measuring interaction force/torque using force/torque sensors mounted on the end-effector, which can, however, only achieve partial force/torque reflection [the reflection of partial interaction force/torque (e.g., that associated with the end-effector) can be guaranteed since it relies on the direct measurement of the interaction via force/torque sensors]. Our study focuses on the full torque-reflecting problem of teleoperators with closed architecture in which case the technical challenge cannot be directly accommodated via incorporating force/torque measurement.

In the context of regulation or tracking control of a single robot with closed architecture, some recent results appear in, e.g., \cite{Wang2017_TAC,Wang2020_TAC} and the basic objective of these results is to achieve the rigorous stabilization/convergence of the tracking or regulation error via dynamic design of the joint velocity (or position) command based on the estimation of the unknown inner controller gains and the unknown parameters of the robot dynamics. From the perspective of systems and control, these results are associated with designing controllers for asymptotically stabilizing the robots with closed architecture [a PD or PID position controller is typically embedded by the robot manufacturers as an inner loop with the gravitational torque compensation (in most cases)].

However, the control objective for teleoperators differs from that for a single robot (e.g., tracking and regulation) in the sense that the closed-loop dynamics of teleoperators are generally (uniformly) marginally stable and of the first kind (see, e.g., \cite{Wang2020b_AUT,Wang2020_AUT}). In addition, as is shown in \cite{Wang2020b_AUT}, manipulability analysis besides the conventional stability analysis needs to be further performed, especially in the context of using dynamic feedback, and in particular, guaranteeing the infinite manipulability with degree one is necessary for facilitating the manipulation of a teleoperator system. However, if the system involves robots with closed architecture (for instance, most industrial/commercial robots that contain an inner PD or PID position controller), possibly due to the proportional and integral action of the inner control loop concerning the position error, its manipulability becomes finite, as is demonstrated in \cite{Wang2018_arXiv} in the case of an inner PD position controller. This gives rise to the consequence that the consensus equilibrium of the teleoperator system involving robots with closed architecture, under the current typical kinematic controllers in the literature (for instance, the synchronization-based kinematic controller in \cite{Wang2018_arXiv} where the joint velocity command is generated directly based on the position difference among the robots), is difficult to be adjusted by a human operator.

This paper provides solutions to the open problem of bilateral control of teleoperators involving robots with closed architecture and subjected to arbitrary bounded time-varying delay (in the sense that the delay can be arbitrarily bounded and piecewise uniformly continuous), by resorting to a new class of dynamic feedback that can guarantee the infinite manipulability of the teleoperator with degree one. Mathematically, the proposed solutions exploit a class of square-integrable functions that hold the possibility of being integral unbounded, based on which, the infinite manipulability of the closed-loop teleoperator is ensured by a new dynamic design of the joint velocity command, facilitating the adjustment of the consensus equilibrium of the teleoperator by a human operator. Specifically, we propose two categories of delay-independent bilateral control for teleoperator systems involving robots with an inner PD or PID position controller: 1) kinematic control and 2) adaptive dynamic control. The kinematic controllers are developed under the guidance of guaranteeing the infinite manipulability of the system with degree one, the separation in \cite{Wang2017_TAC}, and the newly introduced dynamic separation (an extension of the separation in \cite{Wang2017_TAC} for accommodating the context of bilateral teleoperation with enlarged design freedom). The kinematic controller is favorable for its simple implementation, and it can be used in the case of the relatively low requirement concerning the accuracy and dynamic response of the system. For those scenarios that require high accuracy and rapid dynamic response, the proposed adaptive dynamic controller is qualified yet involves increased computational complexity (due to the computation of the required dynamic compensation and parameter adaptation). The dynamic compensation embedded in the joint velocity command is motivated by the result for the control of a single robot with closed architecture in \cite{Wang2020_TAC} (which indicates the necessity and the performance improvement of implementing dynamic compensation in industrial/commercial robots), and the technical difference lies in the control objective and presented new analysis. In particular, our result here concentrates on achieving desired marginally stable dynamics of the first kind (i.e., with the state of its linear time-varying part uniformly converging to a constant vector) and the interaction between the teleoperator system and the human operator/environment while the result in \cite{Wang2020_TAC} considers the tracking and regulation control of a single robot. In addition, with the exploitation of the input-output properties of an inverted form of the dynamics of robot manipulators with closed architecture, we remove the restrictive assumption of uniform exponential stability of a linear time-varying system (due to the adaptation to the gains of the inner PID controller) involved in \cite{Wang2020_TAC}. It is well known that stability of general linear time-varying systems often involves stringent requirements concerning the system coefficient matrix; we here demonstrate by exploiting the input-output properties of inverted dynamics of robots with closed architecture that the boundedness of certain quantities of the closed-loop dynamics can be ensured without relying on such conditions.

\section{Dynamics of Teleoperators with Closed Architecture}

 We take into consideration a master-slave teleoperator system with closed architecture (i.e., an inner PD or PID position control loop is embedded in the master and slave robots). In the case of an inner PID control loop, the dynamics of the teleoperator system can be written as (see, e.g., \cite{Spong2006_Book,Slotine1991_Book,Craig2005_Book,Wang2020_TAC})
 \begin{align}
 \label{eq:1}
 M_1&(q_1)\ddot q_1+C_1(q_1,\dot q_1)\dot q_1+g_1(q_1)\nn\\
 =&-K_{D,1}(\dot q_1-\dot q_{c,1})-K_{P,1}(q_1-q_{c,1})\nn\\
 &-K_{I,1}\int_0^t [q_1(\sigma)-q_{c,1}(\sigma)]d\sigma+\tau_1^\ast
 \\
 \label{eq:2}
 M_2&(q_2)\ddot q_2+C_2(q_2,\dot q_2)\dot q_2+g_2(q_2)\nn\\
 =&-K_{D,2}(\dot q_2-\dot q_{c,2})
 -K_{P,2}(q_2-q_{c,2})\nn\\
 &-K_{I,2}\int_0^t [q_2(\sigma)-q_{c,2}(\sigma)]d\sigma-\tau_2^\ast
 \end{align}
where $q_i\in R^m$ is the joint position, $M_i(q_i)\in R^{m\times m}$ is the inertia matrix, $C_i(q_i,\dot q_i)\in R^{m\times m}$ is the Coriolis and centrifugal matrix, $g_i(q_i)\in R^m$ is the gravitational torque, $q_{c,i}\in R^m$ is the joint position command, $\dot q_{c,i}$ is the joint velocity command, $K_{D,i}$, $K_{P,i}$, and $K_{I,i}$ are the derivative, proportional, and integral gain matrices (diagonal positive definite and unknown), respectively, $i=1,2$, $\tau_1^\ast\in R^m$ is the torque exerted by the human operator on the master (1st robot), and $\tau_2^\ast\in R^m$ is the torque exerted by the slave (2nd robot) on the environment. In the case of an inner PD control loop, $K_{I,1}=K_{I,2}=0$. The particular point that we emphasize here is that in the dynamics given by (\ref{eq:1}) and (\ref{eq:2}), due to the closed architecture, the torque design interface is not available and only the joint position command or velocity command can be specified (as control input). Three well-recognized properties associated with the dynamics (\ref{eq:1}) and (\ref{eq:2}) are listed as follows (see, for instance, \cite{Slotine1991_Book,Spong2006_Book}).

\emph{Property 1:} The inertia matrices $M_1(q_1)$ and $M_2(q_2)$ are symmetric and uniformly positive definite.

\emph{Property 2:} The Coriolis and centrifugal matrices $C_1(q_1,\dot q_1)$ and $C_2(q_2,\dot q_2)$ can be appropriately determined such that $\dot M_1(q_1)-2C_1(q_1,\dot q_1)$ and $\dot M_2(q_2)-2C_2(q_2,\dot q_2)$ are skew-symmetric.

\emph{Property 3:} The dynamics (\ref{eq:1}) and (\ref{eq:2}) depend linearly on constant parameter vectors $\vartheta_1$ and $\vartheta_2$, respectively, and this directly yields
\begin{align}
\label{eq:a1}
M_1(q_1)\dot \zeta_1+C_1(q_1,\dot q_1)\zeta_1+g_1(q_1)=Y_1(q_1,\dot q_1,\zeta_1,\dot \zeta_1)\vartheta_1\\
\label{eq:a2}
M_2(q_2)\dot \zeta_2+C_2(q_2,\dot q_2)\zeta_2+g_2(q_2)=Y_2(q_2,\dot q_2,\zeta_2,\dot \zeta_2)\vartheta_2
\end{align}
where $\zeta_1\in R^m$ and $\zeta_2\in R^m$ are differentiable vectors, $\dot \zeta_1$ and $\dot \zeta_2$ are the derivatives of $\zeta_1$ and $\zeta_2$, respectively, and $Y_1(q_1,\dot q_1,\zeta_1,\dot \zeta_1)$ and $Y_2(q_2,\dot q_2,\zeta_2,\dot \zeta_2)$ are the regressor matrices.

\section{Kinematic Control Using Dynamic Feedback}

In this section, we present a kinematic controller for teleoperators with an inner PID position controller and time-varying communication delay where the nonlinear dynamic effects of teleoperators are assumed to be negligible in the sense of separation in \cite{Wang2017_TAC}. The communication delay is assumed to be piecewise uniformly continuous and bounded. The gravitational torques are assumed to be a priori compensated, namely the assumption that $g_1(q_1)=g_2(q_2)=0$ holds in the teleoperator system given by (\ref{eq:1}) and (\ref{eq:2}), and this assumption shall be very mild since this practice is commonly employed in most industrial/commercial robots (for instance, the UR10 robot manufactured by Universal Robots). Specifically, in the case of control of teleoperators with closed architecture, we would typically rely on the velocity control mode (e.g., the UR10 robot), similar to the case of continuous path control of robots with closed architecture. In this case, we may notice an important point that the joint position commands $q_{c,1}$ and $q_{c,2}$ in the inner control loop cannot be accurately specified from the outer loop since the outer loop can only specify the joint velocity command; a similar case is reflected in the integral action. Let $\dot q_{c,i}^\ast$ and $q_{c,i}^\ast$ denote the joint velocity and position commands that are specified or designed by the outer loop, respectively, $\forall i=1,2$, and then the following relation holds
\begin{align}
q_{c,i}^\ast=&q_{c,i}+\theta_i\\
\dot q_{c,i}^\ast=&\dot q_{c,i}
\end{align}
with $\theta_i$ being an unknown constant or slowly time-varying vector, $\forall i=1,2$. This implies that the joint position command used in the inner control loop generally differs from the one generated in the outer loop (i.e., specified by the designer or user) with an unknown constant or slowly time-varying offset.

With the availability of the joint position command of the inner control loop, we design the joint velocity commands $\dot q_{c,i}^\ast$, $i=1,2$ as
\begin{align}
\label{eq:7}
\dot q_{c,1}^\ast=&-\lambda[q_1-q_2(t-T_2)]+\lambda_P(q_1-q_{c,1})\nn\\
&+\lambda_{\mathcal M}\int_0^t [q_1(\sigma)-q_{c,1}(\sigma)]d\sigma\\
\label{eq:8}
\dot q_{c,2}^\ast=&-\lambda[q_2-q_1(t-T_1)]+\lambda_P(q_2-q_{c,2})\nn\\
&+\lambda_{\mathcal M}\int_0^t [q_2(\sigma)-q_{c,2}(\sigma)]d\sigma
\end{align}
where $\lambda$, $\lambda_P$, and $\lambda_{\mathcal M}$ are positive design constants, and $T_1$ and $T_2$ are time-varying forward and backward delays (piecewise uniformly continuous and bounded), respectively. The distinctive point is the incorporation of $\lambda_{\mathcal M}\int_0^t [q_i(\sigma)-q_{c,i}(\sigma)]d\sigma$, $i=1,2$ in the definition of the joint velocity commands. This action, on the one hand, renders the proposed controller to rely on dynamic feedback, and on the other hand, will be shown to be indispensable for ensuring the easy maneuvering or more precisely the infinite manipulability of the teleoperator system. This practice is partially motivated by the result in \cite{Wang2018_CAC}, but we might note the apparently different scenario considered here, i.e., the teleoperator system involves robots with closed architecture and no torque design interface is available.

\emph{Remark 1:} The proposed kinematic control relies on the availability of the joint position commands of the inner control loop.
 This suggests the necessity of the accessibility (to the outer loop) of the joint position command generated by the inner control loop in the joint velocity control mode (which is set by robot manufacturers); there is neither safety nor conflicting issue and only the potential expansion or enhancement of robots in relevant application scenarios (for instance, bilateral teleoperation). It seems favorable that the joint position command generated in the inner control loop can be accessed in robots manufactured by Universal Robots (for instance, the UR10 robot). If this joint position command is unavailable, the joint velocity commands for the master and slave robots can be specified as
 \begin{align}
\label{eq:9}
\dot q_{c,1}^\ast=&-\lambda[q_1-q_2(t-T_2)]+\lambda_P(q_1-q_{c,1}^\ast)\nn\\
&+\lambda_{\mathcal M}\int_0^t [q_1(\sigma)-q_{c,1}^\ast(\sigma)]d\sigma\\
\label{eq:10}
\dot q_{c,2}^\ast=&-\lambda[q_2-q_1(t-T_1)]+\lambda_P(q_2-q_{c,2}^\ast)\nn\\
&+\lambda_{\mathcal M}\int_0^t [q_2(\sigma)-q_{c,2}^\ast(\sigma)]d\sigma.
\end{align}
However, under the action of this control, the unknown offset between the joint
position command generated in the outer loop and the
one generated in the inner loop would give rise to
the synchronization error and also influence the manipulability
of the teleoperator system.

With the joint velocity command being generated by (\ref{eq:7}) and (\ref{eq:8}), the closed-loop dynamics of the teleoperator can be written as
\begin{align}
\label{eq:11}
\dot q=&\mathcal F_D(q)+\dot \psi+\lambda_P\psi+\lambda_{\mathcal M}\int_0^t \psi(\sigma)d\sigma
\end{align}
where $q=[q_1^T,q_2^T]^T$, $\psi=[\psi_1^T,\psi_2^T]^T$ with $\psi_i=q_i-q_{c,i}$, $i=1,2$, and $\mathcal F_D(q)=-\lambda\begin{bmatrix}q_1-q_2(t-T_2)\\ q_2-q_1(t-T_1)\end{bmatrix}.$

\emph{Theorem 1:} Suppose that the inner PID control can guarantee that the joint velocity and position tracking errors and the integral of the joint position tracking error are square-integrable and bounded in free motion\footnote{See Remark 2 for the discussion concerning this assumption.}. The kinematic controller given by (\ref{eq:7}) and (\ref{eq:8}) for the teleoperator system given by (\ref{eq:1}) and (\ref{eq:2}) with the gravitational torques being exactly compensated a priori ensures the position synchronization of the master and slave robots in free motion and static torque quasi-reflection. In addition, the manipulability of the teleoperator system is infinite with degree one.

\emph{Proof:} We first consider the case of free motion. According to the assumption, $\dot \psi\in \mathcal L_2\cap\mathcal L_\infty$, $\psi\in\mathcal L_2\cap\mathcal L_\infty$, and $\int_0^t \psi(\sigma)d\sigma\in\mathcal L_2\cap\mathcal L_\infty$ (this is guaranteed by the inner control of the master and slave robots). The state $q$ of the system (\ref{eq:11}) with $\dot\psi+\lambda_P\psi+\lambda_{\mathcal M}\int_0^t \psi(\sigma)d\sigma=0$ uniformly converge to a constant vector in accordance with \cite{Munz2011b_TAC}, and hence this system is uniformly marginally stable and of the first kind (see \cite{Wang2020_AUT}). For the system (\ref{eq:11}) with $q$ as the state, the application of \cite[Proposition~3]{Wang2020_AUT} yields the result that $\dot q\in\mathcal L_2\cap\mathcal L_\infty$. This leads us to immediately obtain that $\mathcal F_D(q)\in\mathcal L_2\cap\mathcal L_\infty$. From (\ref{eq:1}) and (\ref{eq:2}) and using Property 1, we have that $\ddot q_i\in\mathcal L_\infty$, implying that $\dot q_i$ is uniformly continuous, $\forall i=1,2$. The result that $\psi\in\mathcal L_\infty$, $\dot \psi\in\mathcal L_\infty$, and $\dot q\in\mathcal L_\infty$ implies that $\int_0^t\psi(\sigma)d\sigma$, $\psi$, and $q$ are uniformly continuous. From (\ref{eq:9}) and (\ref{eq:10}), we have that $\dot q_{c,i}=\dot q_{c,i}^\ast$ is piecewise uniformly continuous, $\forall i=1,2$. Hence, $\dot \psi$ is piecewise uniformly continuous. Using the properties of square-integrable and uniformly continuous functions \cite[p.~232]{Desoer1975_Book} yields the result that $\int_0^t\psi(\sigma)d\sigma\to 0$ and $\psi\to 0$ as $t\to\infty$. From the standard generalized Barbalat's lemma, we have that $\dot \psi\to 0$ as $t\to\infty$. From \cite[Proposition~3]{Wang2020_AUT}, we have that $\dot q\to 0$ as $t\to\infty$ and hence that $\mathcal F_D(q)\to 0$ as $t\to\infty$. With a typical procedure, we obtain that $q_1-q_2(t-T_2)=q_1-q_2+\int_0^{T_2}\dot q_2(t-\sigma)d\sigma\to q_1-q_2\to 0$ and $q_2-q_1(t-T_1)=q_2-q_1+\int_0^{T_1}\dot q_1(t-\sigma)d\sigma\to q_2-q_1\to 0$ as $t\to\infty$.

Now consider the case of contact motion, and we follow the standard procedure to analyze the torque reflection property of the teleoperator system (see, e.g., \cite{Lee2006_TRO,Liu2013_AUT}). Specifically, as $\dot q_i$, $\ddot q_i$, and $\dot q_{c,i}$ (or $\dot q_{c,i}^\ast$) converge to zero, $\forall i=1,2$, it can be shown that
\begin{align}
\tau_1^\ast&\to \frac{\lambda}{\lambda_{\mathcal M}}K_{I,1}(q_1-q_2)\\
\tau_2^\ast&\to -\frac{\lambda}{\lambda_{\mathcal M}}K_{I,2}(q_2-q_1)
\end{align}
which yields the result that $\tau_1^\ast \to K_{I,1}K_{I,2}^{-1}\tau_2^\ast$ where $K_{I,1}K_{I,2}^{-1}$ is diagonal positive definite. This is referred to as static torque quasi-reflection, similar to the static torque reflection in the sense of \cite{Lee2006_TRO}.

We next analyze the manipulability of the teleoperator system using some standard tools concerning the input-output stability of dynamical systems (see, e.g., \cite{Desoer1975_Book,Schaft2000_Book,Khalil2002_Book}). The $\mathcal L_2$-gain from $\tau_1^\ast$ to $\dot \psi_1$, that from $\tau_1^\ast$ to $\psi_1$, and that from $\tau_1^\ast$ to $\int_0^t \psi_1(\sigma)d\sigma$ can be considered to be finite under the standard assumption in Theorem 1. Denote by $c_1$ the upper bound of the $\mathcal L_2$-gain from $\tau_1^\ast$ to $\int_0^t\psi_1(\sigma)d\sigma$. From (\ref{eq:11}), the $\mathcal L_2$-gain from $\int_0^t\psi(\sigma)d\sigma$ to $\dot q$ is less than or equal to $c_1^\ast\sup_\omega\sqrt{(\lambda_{\mathcal M}-\omega^2)^2+\lambda_P^2\omega^2}$ with $c_1^\ast$ being certain positive constant, by following the standard practice and using the result concerning the system (\ref{eq:11}) shown before, i.e., $\dot q\in\mathcal L_2$ in the case that $\dot \psi+\lambda_P\psi+\lambda_{\mathcal M}\int_0^t \psi(\sigma)d\sigma\in\mathcal L_2$. Let $q_\text{ave}=(1/2)(q_1+q_2)$ and $\psi_\text{ave}=(1/2)(\psi_1+\psi_2)$, and we then obtain that the $\mathcal L_2$-gain from $\psi_\text{ave}$ to $\dot q_\text{ave}$ is also less than or equal to $c_1^\ast\sup_\omega\sqrt{(\lambda_{\mathcal M}-\omega^2)^2+\lambda_P^2\omega^2}$. The $\mathcal L_2$-gain from $\dot q_\text{ave}$ to $q_\text{ave}-q_\text{ave}(0)$ can be obtained by calculating the standard $\mathcal H_\infty$ norm as $\sup_\omega({1}/{|\omega|})$. This leads us to obtain the $\mathcal L_2$-gain from $\tau_1^\ast$ to $q_\text{ave}-q_\text{ave}(0)$ as
\begin{align}
&{\mathcal M}_{\tau_1^\ast\mapsto q_\text{ave}-q_\text{ave}(0)}\nn\\
&\le \frac{c_1 c_1^\ast}{2}\sup_\omega \sqrt{\frac{\lambda_{\mathcal M}^2+(\lambda_P^2-2\lambda_{\mathcal M})\omega^2+\omega^4}{\omega^2}}= \infty.
\end{align}
From the typical practice of calculating the gains for nonlinear dynamical systems, ${\mathcal M}_{\tau_1^\ast\mapsto q_\text{ave}-q_\text{ave}(0)}$ has the same order as its upper bound. Therefore, the manipulability of the system is infinite. On the other hand, the mapping from $\tau_1^\ast$ to $q_\text{ave}-q_\text{ave}(0)$ involves only one pure integral operation, and thus, the manipulability of the system is infinite with degree one. The infinite manipulability of the system under the action of $-\tau_2^\ast$ (namely, the torque exerted by the environment on the slave) can be similarly demonstrated. \hfill {\small$\blacksquare$}

\emph{Remark 2:} The kinematic controller developed for teleoperators here exploits the properties of the separation approach in \cite{Wang2017_TAC}. In \cite{Wang2017_TAC}, the square-integrability and boundedness of the joint velocity tracking error is shown to be necessary for ensuring the stability of the kinematic controllers for task-space tracking of a single robot with closed architecture. In the context of bilateral control for teleoperators with an inner PID position controller, we mainly exploit the integral unboundedness of the integral of the joint position tracking error (i.e., $\int_0^t [q_i(\sigma)-q_{c,i}(\sigma)]d\sigma$, $\forall i=1,2$) in addition to the square-integrability of the joint velocity tracking error and that of the joint position tracking error and its integral (the properties of such square-integrable functions are systematically demonstrated to be important for facilitating the human-system interaction in \cite{Wang2020b_AUT}). %The square-integrability of the joint position and velocity tracking errors, under a PD position controller, holds in the practical sense, especially in the case of relatively slow motion (see, e.g., \cite[Remark~7]{Wang2018_TCST}).

\section{Adaptive Dynamic Control}

As is previously discussed, the stability of kinematic control relies on the assumption that the inner joint control loop ensures the square-integrability and boundedness of the joint velocity tracking error, joint position tracking error, and integral of the joint position error. This assumption tends to become ad hoc (in the rigorous sense) in the case of fast motion (see, e.g., \cite{Wang2020_TAC}). To rigorously take into consideration the dynamic effects of the teleoperator system, we define two vectors $z_1$ and $z_2$ with the dynamics given as
\begin{align}
\label{eq:21}
\ddot z_1=&-\alpha \dot q_1-\lambda[\xi_1-\xi_2(t-T_2)]\nn\\
&+\Lambda_{D,1}(\dot q_1-\dot z_1)+\Lambda_{P,1}(q_1-z_1)\nn\\
&+\Lambda_{I,1}\int_0^t [q_1(\sigma)-z_1(\sigma)]d\sigma\\
\label{eq:22}
\ddot z_2=&-\alpha \dot q_2-\lambda[\xi_2-\xi_1(t-T_1)]\nn\\
&+\Lambda_{D,2}(\dot q_2-\dot z_2)+\Lambda_{P,2}(q_2-z_2)\nn\\
&+\Lambda_{I,2}\int_0^t [q_2(\sigma)-z_2(\sigma)]d\sigma
\end{align}
where $\alpha$ is a positive design constant, $\xi_1$ and $\xi_2$ are defined as (the same as \cite{Chopra2006,Chopra2008})
\begin{align}
\label{eq:23}
\xi_1=&\dot q_1+\alpha q_1\\
\label{eq:24}
\xi_2=&\dot q_2+\alpha q_2,
\end{align}
 and $\Lambda_{D,i}$, $\Lambda_{P,i}$, and $\Lambda_{I,i}$ are diagonal positive definite matrices, $\forall i=1,2$. Define
 \begin{align}
 \label{eq:25}
\psi_{1}^\ast=&q_1-z_1\\
\label{eq:26}
\psi_2^\ast=&q_2-z_2\\
\label{eq:27}
\zeta_1^\ast=&\dot z_1-\gamma_1\psi_1^\ast-\gamma_1^\ast\int_0^t \psi_1^\ast(\sigma)d\sigma\\
\label{eq:28}
\zeta_2^\ast=&\dot z_2-\gamma_2\psi_2^\ast-\gamma_2^\ast\int_0^t \psi_2^\ast(\sigma)d\sigma\\
\label{eq:29}
s_1=&\dot q_1-\zeta_1^\ast\\
\label{eq:30}
s_2=&\dot q_2-\zeta_2^\ast
\end{align}
where $\gamma_i$ and $\gamma_i^\ast$ are positive design constants, $\forall i=1,2$. The joint velocity commands for the teleoperator system are given as
\begin{align}
\label{eq:31}
\dot q_{c,1}=&\dot z_1-\hat{\mathcal K}_{P,1} (q_{c,1}-z_1)-\hat{\mathcal K}_{I,1}\int_0^t [q_{c,1}(\sigma)-z_1(\sigma)]d\sigma\nn\\
&+{\rm diag}[\hat w_1]Y_1(q_1,\dot q_1,\zeta_1^\ast,\dot \zeta_1^\ast)\hat\vartheta_1\\
\label{eq:32}
\dot q_{c,2}=&\dot z_2-\hat{\mathcal K}_{P,2} (q_{c,2}-z_2)-\hat{\mathcal K}_{I,2}\int_0^t [q_{c,2}(\sigma)-z_2(\sigma)]d\sigma\nn\\
&+{\rm diag}[\hat w_2]Y_2(q_2,\dot q_2,\zeta_2^\ast,\dot \zeta_2^\ast)\hat\vartheta_2
\end{align}
where $\hat{\mathcal K}_{P,i}={\rm diag}[\hat w_{P,i}]$ and $\hat {\mathcal K}_{I,i}={\rm diag}[\hat w_{I,i}]$ with $\hat w_{P,i}$ and $\hat w_{I,i}$ being the estimates of $m$-dimensional vectors $w_{P,i}$ and $w_{I,i}$, respectively, which satisfy the property that $K_{D,i}^{-1}K_{P,i}={\rm diag}[w_{P,i}]$ and $K_{D,i}^{-1}K_{I,i}={\rm diag}[w_{I,i}]$, $\hat\vartheta_i$ is the estimate of $\vartheta_i$, and $\hat w_i$ is the estimate of a $m$-dimensional vector $w_i$ where the $k$-th component of $w_i$ is the $k$-th diagonal entry of $K_{D,i}^{-1}$, $k=1,\dots,m$, $\forall i=1,2$. The adaptation laws for $\hat\vartheta_1$, $\hat \vartheta_2$, $\hat w_1$, $\hat w_2$, $\hat w_{P,1}$, and $\hat w_{P,2}$ are given as
\begin{align}
\label{eq:33}
\dot{\hat \vartheta}_1=&-\Gamma_1 Y_1^T(q_1,\dot q_1,\zeta_1^\ast,\dot \zeta_1^\ast) s_1\\
\label{eq:34}
\dot{\hat \vartheta}_2=&-\Gamma_2 Y_2^T(q_2,\dot q_2,\zeta_2^\ast,\dot \zeta_2^\ast) s_2\\
\label{eq:35}
\dot{\hat w}_1=&-\Gamma_1^\ast {\rm diag}[Y_1(q_1,\dot q_1,\zeta_1^\ast,\dot \zeta_1^\ast)\hat\vartheta_1]s_1\\
\label{eq:36}
\dot{\hat w}_2=&-\Gamma_2^\ast {\rm diag}[Y_2(q_2,\dot q_2,\zeta_2^\ast,\dot \zeta_2^\ast)\hat\vartheta_2]s_2\\
\label{eq:37}
\dot{\hat w}_{P,1}=&-\Gamma^\ast_{P,1}{\rm diag}[z_1-q_{c,1}]s_1\\
\label{eq:38}
\dot{\hat w}_{P,2}=&-\Gamma^\ast_{P,2}{\rm diag}[z_2-q_{c,2}]s_2\\
\label{eq:39}
\dot{\hat w}_{I,1}=&-\Gamma^\ast_{I,1}{\rm diag}\Big[\int_0^t[ z_1(\sigma)-q_{c,1}(\sigma)]d\sigma\Big] s_1\\
\label{eq:40}
\dot{\hat w}_{I,2}=&-\Gamma^\ast_{I,2}{\rm diag}\Big[\int_0^t[ z_2(\sigma)-q_{c,2}(\sigma)]d\sigma\Big] s_2
\end{align}
where $\Gamma_1$ and $\Gamma_2$ are symmetric positive definite matrices, and $\Gamma_1^\ast$, $\Gamma_2^\ast$, $\Gamma^\ast_{P,1}$,  $\Gamma^\ast_{P,2}$, $\Gamma^\ast_{I,1}$, and $\Gamma^\ast_{I,2}$ are diagonal positive definite matrices. The dynamics of the teleoperator system can be described by the following dynamic-cascaded system
\be
\label{eq:41}
\begin{cases}
\dot \xi_1=-\lambda[\xi_1-\xi_2(t-T_2)]+\ddot\psi_1^\ast\\
\qquad+\Lambda_{D,1}\dot \psi_1^\ast+\Lambda_{P,1} \psi^\ast_1+\Lambda_{I,1}\int_0^t\psi_1^\ast(\sigma)d\sigma\\
\dot \xi_2=-\lambda[\xi_2-\xi_1(t-T_1)]+\ddot\psi_2^\ast\\
\qquad+\Lambda_{D,2}\dot \psi_2^\ast+\Lambda_{P,2} \psi^\ast_2+\Lambda_{I,2}\int_0^t\psi_2^\ast(\sigma)d\sigma\\
M_1(q_1)\dot s_1+C_1(q_1,\dot q_1)s_1\\
=-K_{D,1} [\dot \psi_1^\ast+{\mathcal K}_{P,1}\psi_1^\ast+{\mathcal K}_{I,1}\int_0^t \psi_1^\ast(\sigma)d\sigma]\\
\quad+K_{D,1}{\rm diag}[z_1-q_{c,1}]\Delta w_{P,1}\\
\quad+K_{D,1}{\rm diag}[\int_0^t [z_1(\sigma)-q_{c,1}(\sigma)]d\sigma]\Delta w_{I,1}\\
\quad+Y_1(q_1,\dot q_1,\zeta_1^\ast,\dot\zeta_1^\ast)\Delta\vartheta_1\\
\quad+K_{D,1}{\rm diag}[Y_1(q_1,\dot q_1,\zeta_1^\ast,\dot \zeta_1^\ast)\hat\vartheta_1]\Delta w_1+\tau_1^\ast\\
M_2(q_2)\dot s_2+C_2(q_2,\dot q_2)s_2\\
=-K_{D,2} [\dot \psi_2^\ast+{\mathcal K}_{P,2}\psi_2^\ast+{\mathcal K}_{I,2}\int_0^t \psi_2^\ast(\sigma)d\sigma]\\
\quad+K_{D,2}{\rm diag}[z_2-q_{c,2}]\Delta w_{P,2}\\
\quad+K_{D,2}{\rm diag}[\int_0^t [z_2(\sigma)-q_{c,2}(\sigma)]d\sigma]\Delta w_{I,2}\\
\quad+Y_2(q_2,\dot q_2,\zeta_2^\ast,\dot \zeta_2^\ast)\Delta\vartheta_2\\
\quad+K_{D,2}{\rm diag}[Y_2(q_2,\dot q_2,\zeta_2^\ast,\dot \zeta_2^\ast)\hat\vartheta_2]\Delta w_2-\tau_2^\ast
\end{cases}
\ee
and the adaptation laws (\ref{eq:33})-(\ref{eq:40}), where $\mathcal K_{P,i}={\rm diag}[w_{P,i}]$, $\mathcal K_{I,i}={\rm diag}[w_{I,i}]$, $\Delta\vartheta_i=\hat\vartheta_i-\vartheta_i$, $\Delta w_i=\hat w_i-w_i$, $\Delta w_{P,i}=\hat w_{P,i}-w_{P,i}$, and $\Delta w_{I,i}=\hat w_{I,i}-w_{I,i}$, $i=1,2$.

\emph{Theorem 2:} Suppose that $\gamma_i$ and $\gamma_i^\ast$ are chosen such that
\begin{align}
\label{eq:42}
&\gamma_i^\ast I_m\ge K_{D,i}^{-1} K_{I,i}\\
\label{eq:43}
& \gamma_i I_m \ge\epsilon I_m+(\gamma_i^\ast I_m-K_{D,i}^{-1}K_{I,i})K_{D,i}K_{P,i}^{-1}
\end{align}
where $I_m$ is the $m\times m$ identity matrix, and $\epsilon$ is a positive constant that satisfies the property that $\epsilon I_m\le K_{D,i}^{-1}K_{P,i}$ and can be arbitrarily small, $\forall i=1,2$. Then, the adaptive controller given by (\ref{eq:31}), (\ref{eq:32}), (\ref{eq:33})-(\ref{eq:40}) with $z_1$ and $z_2$ being given by (\ref{eq:21}) and (\ref{eq:22}), respectively and $\zeta_1^\ast$ and $\zeta_2^\ast$ given by (\ref{eq:27}) and (\ref{eq:28}), respectively for the teleoperator system given by (\ref{eq:1}) and (\ref{eq:2}) ensures the position synchronization of the master and slave robots in free motion. If the gravitational torques are a priori compensated, the adaptive controller ensures static torque quasi-reflection provided that $\Lambda_{D,i}$, $\Lambda_{P,i}$, and $\Lambda_{I,i}$ are chosen such that the linear system
\be
\label{eq:a3}
\frac{d^3 y_i}{dt^3}+\Lambda_{D,i}\frac{d^2 y_i}{dt^2}+\Lambda_{P,i}\frac{d y_i}{dt}+\Lambda_{I,i}y_i=0
 \ee
 with $y_i\in R^m$ is exponentially stable, $\forall i=1,2$, and the teleoperator system is infinitely manipulable with degree one.

\emph{Proof:} We first take into consideration the case of free motion, i.e., $\tau_1^\ast=\tau_2^\ast=0$. Considering the dynamic-cascaded property of the
closed-loop system, we can first analyze the system given by
the third and fourth subsystems of (\ref{eq:41}) and (\ref{eq:33})-(\ref{eq:40}). In particular, consider the following function
\begin{align}
V_i=&\frac{1}{2}s_i^T M_i(q_i)s_i+\frac{1}{2}\Delta\vartheta_i^T \Gamma_i^{-1}\Delta\vartheta_i+\frac{1}{2}\Delta w_i^T K_{D,i}\Gamma_i^{\ast-1}\Delta w_i\nn\\
&+\frac{1}{2}\Delta w_{P,i}^T K_{D,i}\Gamma_{P,i}^{\ast-1}\Delta w_{P,i}+\frac{1}{2}\Delta w_{I,i}^T K_{D,i}\Gamma_{I,i}^{\ast-1}\Delta w_{I,i}
\end{align}
and its derivative along the trajectories of the system can be written as (using Property 2)
\begin{align}
\dot V_i=&-\left[\dot \psi_i^\ast+\gamma_i \psi_i^\ast+\gamma_i^\ast \int_0^t \psi_i^\ast(\sigma)d\sigma\right]^T\nn\\
&\times\left[K_{D,i}\dot \psi_i^\ast+K_{P,i} \psi_i^\ast+K_{I,i}\int_0^t \psi_i^\ast(\sigma)d\sigma\right]\nn\\
=&-\left[\dot \psi_i^\ast+\epsilon \psi_i^\ast+K_{D,i}^{-1}K_{I,i}\int_0^t \psi_i^\ast(\sigma)d\sigma\right.\nn\\
&\left.\quad+(\gamma_i-\epsilon)\psi_i^\ast+(\gamma_i^\ast I_m-K_{D,i}^{-1}K_{I,i})\int_0^t \psi_i^\ast(\sigma)d\sigma\right]^T\nn\\
&\times K_{D,i}\left[\dot \psi_i^\ast+\epsilon\psi_i^\ast+K_{D,i}^{-1}K_{I,i}\int_0^t \psi_i^\ast(\sigma)d\sigma\right.\nn\\
&\left.\quad+(K_{D,i}^{-1}K_{P,i}-\epsilon I_m)\psi_i^\ast\right]\nn\\
=&-\left[\dot \psi_i^\ast+\epsilon \psi_i^\ast+K_{D,i}^{-1}K_{I,i}\int_0^t \psi_i^\ast(\sigma)d\sigma\right.\nn\\
&\left.\quad+[\gamma_i I_m-\epsilon I_m-(\gamma_i^\ast I_m-K_{D,i}^{-1}K_{I,i})K_{D,i}K_{P,i}^{-1}]\psi_i^\ast\right.\nn\\
&\left.\quad +(\gamma_i^\ast I_m-K_{D,i}^{-1}K_{I,i})K_{D,i}K_{P,i}^{-1}\right.\nn\\
&\left.\quad\times\Big(\psi_i^\ast+K_{D,i}^{-1}K_{P,i}\int_0^t \psi_i^\ast(\sigma)d\sigma\Big)\right]^T\nn\\
&\quad\times  K_{D,i}\left[\dot \psi_i^\ast+\epsilon\psi_i^\ast+K_{D,i}^{-1}K_{I,i}\int_0^t \psi_i^\ast(\sigma)d\sigma\right.\nn\\
&\left.\quad+(K_{D,i}^{-1}K_{P,i}-\epsilon I_m)\psi_i^\ast\right].
\end{align}
Next, we consider the function
\begin{align}
V_i^{\ast}=&\frac{1}{2}\psi_i^{\ast T}{\mathcal D}_{1,i}\psi_i^\ast+\frac{1}{2}\int_0^t \psi_i^{\ast T} (\sigma)d\sigma{\mathcal D}_{2,i} \int_0^t \psi_i^{\ast} (\sigma)d\sigma\nn\\
&+\frac{1}{2}\left[\psi_i^\ast+K_{D,i}^{-1}K_{P,i}\int_0^t \psi_i^\ast(\sigma)d\sigma\right]^T {\mathcal D}_{3,i} \nn\\
&\times\left[\psi_i^\ast+K_{D,i}^{-1}K_{P,i}\int_0^t \psi_i^\ast(\sigma)d\sigma\right]
\end{align}
with $\mathcal D_{1,i}$, $\mathcal D_{2,i}$, and $\mathcal D_{3,i}$ being given by
\begin{align}
{\mathcal D}_{1,i}=&K_{P,i}-\epsilon K_{D,i}+(\gamma_i I_m-\epsilon I_m-\gamma_i ^\ast K_{D,i}K_{P,i}^{-1}\nn\\
&+K_{P,i}^{-1}K_{I,i})K_{D,i}\\
{\mathcal D}_{2,i}=&K_{D,i}^{-1}K_{I,i}(K_{P,i}-\epsilon K_{D,i})+(\gamma_i I_m-\epsilon I_m\nn\\
&-\gamma_i ^\ast K_{D,i}K_{P,i}^{-1} +K_{P,i}^{-1}K_{I,i})K_{I,i}\nn\\
&+(\gamma_i^\ast I_m-K_{D,i}^{-1}K_{I,i})K_{D,i}K_{P,i}^{-1}K_{I,i}\\
{\mathcal D}_{3,i}=&(\gamma_i^\ast K_{D,i}-K_{I,i})K_{D,i}K_{P,i}^{-1}
\end{align}
which are all diagonal nonnegative definite, and hence $V_i^{\ast}$ is nonnegative, $\forall i=1,2$. Then, taking $V_i+V_i^{\ast}$ as the Lyapunov-like function candidate, we have that
\begin{align}
\dot V_i&+\dot V_i^{\ast}\nn\\
=&-\left[\dot \psi_i^\ast+\epsilon \psi_i^\ast+K_{D,i}^{-1}K_{I,i}\int_0^t \psi_i^\ast(\sigma)d\sigma\right]^T\nn\\
 &\times K_{D,i}\left[\dot \psi_i^\ast+\epsilon \psi_i^\ast+K_{D,i}^{-1}K_{I,i}\int_0^t \psi_i^\ast(\sigma)d\sigma\right]\nn\\
&-\epsilon \psi_i^{\ast T}(K_{P,i}-\epsilon K_{D,i})\psi_i^\ast-\psi_i^{\ast T}[\gamma_i I_m-\epsilon I_m\nn\\
&-(\gamma_i^\ast I_m-K_{D,i}^{-1}K_{I,i})K_{D,i}K_{P,i}^{-1}]K_{P,i}\psi_i^\ast\nn\\
&-\left[\int_0^t \psi_i^{\ast }(\sigma)d\sigma\right]^T (\gamma_i^\ast I_m-K_{D,i}^{-1}K_{I,i})\nn\\
&\times K_{I,i}\int_0^t \psi_i^{\ast}(\sigma)d\sigma \le 0, \forall i=1,2.
\end{align}
This implies that $s_i\in\mathcal L_\infty$, $\dot \psi_i^\ast+\epsilon \psi_i^\ast+K_{D,i}^{-1}K_{I,i}\int_0^t \psi_i^\ast(\sigma)d\sigma\in\mathcal L_2$, $\hat\vartheta_i\in\mathcal L_\infty$, $\hat w_i\in\mathcal  L_\infty$, $\hat w_{P,i}\in\mathcal L_\infty$, and $\hat w_{I,i}\in\mathcal L_\infty$, $\forall i=1,2$. From the input-output properties of exponentially stable and strictly proper linear systems \cite[p.~59]{Desoer1975_Book}, we have that  $\dot\psi_i^\ast\in\mathcal L_2\cap\mathcal L_\infty$, $\psi_i^\ast\in\mathcal L_2\cap\mathcal L_\infty$, and $\int_0^t\psi_i^\ast(\sigma)d\sigma\in\mathcal L_2\cap\mathcal L_\infty$, $\forall i=1,2$. The first two subsystems of (\ref{eq:41}) can be rewritten as
\begin{align}
\label{eq:49}
\dot \xi=\mathcal F_D(\xi)+\ddot \psi^\ast+\Lambda_D\dot \psi^\ast+\Lambda_P\psi^\ast+\Lambda_I\int_0^t \psi^\ast(\sigma)d\sigma
\end{align}
where $\xi=[\xi_1^T,\xi_2^T]^T$, $\psi^\ast=[\psi_1^{\ast T},\psi_2^{\ast T}]^T$, $\Lambda_D={\rm diag}[\Lambda_{D,1},\Lambda_{D,2}]$, $\Lambda_P={\rm diag}[\Lambda_{P,1},\Lambda_{P,2}]$, and $\Lambda_I={\rm diag}[\Lambda_{I,1},\Lambda_{I,2}]$. Following \cite{Wang2020_AUT,Wang2019_ACC}, we rewrite (\ref{eq:49}) as
\begin{align}
\label{eq:52}
\frac{d}{dt}(\xi-\dot\psi^\ast)=&\mathcal F_D(\xi-\dot\psi^\ast)+u^\ast
\end{align}
where $u^\ast=\mathcal F_D(\dot \psi^\ast)+\Lambda_D\dot \psi^\ast+\Lambda_P\psi^\ast+\Lambda_I\int_0^t \psi^\ast(\sigma)d\sigma$. The system (\ref{eq:52}) with $\xi-\dot\psi^\ast$ as the state and $u^\ast=0$ is uniformly marginally stable and of the first kind (see \cite{Wang2020_AUT}) in accordance with \cite{Munz2011b_TAC} and the standard linear system theory. Using the input-output properties of uniformly marginally stable linear systems of the first kind \cite[Proposition~3]{Wang2020_AUT} yields the result that $\dot\xi-\ddot\psi^\ast\in\mathcal L_\infty$ since $u^\ast\in\mathcal L_\infty$. Hence, $\mathcal F_D(\xi-\dot\psi^\ast)\in\mathcal L_\infty$ and $\mathcal F_D(\xi)\in\mathcal L_\infty$. Equations (\ref{eq:21}) and (\ref{eq:22}) can be rewritten as
\begin{align}
\label{eq:53}
\ddot z=&-\alpha \dot z+\mathcal F_D(\xi)\nn\\
&+\Lambda_D\dot \psi^\ast+\Lambda_P\psi^\ast+\Lambda_I\int_0^t \psi^\ast(\sigma)d\sigma-\alpha \dot \psi^\ast
\end{align}
with $z=[z_1^T,z_2^T]^T$. Using the input-output properties of exponentially stable and strictly proper linear systems \cite[p.~59]{Desoer1975_Book}, we have from (\ref{eq:53}) that $\dot z\in\mathcal L_\infty$ and $\ddot z\in\mathcal L_\infty$. From (\ref{eq:27}) and (\ref{eq:28}), we have that $\zeta_i^\ast\in\mathcal L_\infty$, $\forall i=1,2$. Hence, $\dot q_i=s_i+\zeta_i^\ast\in\mathcal L_\infty$, $\forall i=1,2$. The dynamics (\ref{eq:1}) and (\ref{eq:2}) can be reformulated as an inverted form given as
\begin{align}
\label{eq:54}
&K_{D,i}(\dot q_i-\dot q_{c,i})\nn\\
&=-K_{P,i}(q_i-q_{c,i})-K_{I,i}\int_0^t [q_i(\sigma)-q_{c,i}(\sigma)]d\sigma\nn\\
&\quad\text{ }\underbrace{-\frac{d}{dt}[M_i(q_i)\dot q_i]}_{u_{i,1}}+\underbrace{\dot M_i(q_i)\dot q_i-C_i(q_i,\dot q_i)\dot q_i-g_i(q_i)}_{u_{i,2}},
\end{align}
$i=1,2$. The system (\ref{eq:54}) with $q_i-q_{c,i}$ and $\int_0^t [q_i(\sigma)-q_{c,i}(\sigma)]d\sigma$ as the output and with $u_{i,1}=-\frac{d}{dt}[M_i(q_i)\dot q_i]=0$ and $u_{i,2}=\dot M_i(q_i)\dot q_i-C_i(q_i,\dot q_i)\dot q_i-g_i(q_i)=0$ is exponentially stable and strictly proper from the standard linear system theory, $\forall i=1,2$. Note that $M_i(q_i)\dot q_i\in\mathcal L_\infty$ and $\dot M_i(q_i)\dot q_i-C_i(q_i,\dot q_i)\dot q_i-g_i(q_i)\in\mathcal L_\infty$, $\forall i=1,2$. Hence, $\int_0^t u_{i,1}(\sigma)d\sigma\in\mathcal L_\infty$ and $u_{i,2}\in\mathcal L_\infty$, $\forall i=1,2$. From \cite{Wang2020_AUT}, the part of the output corresponding to $u_{i,1}$ is bounded, and from the input-output properties of linear systems \cite[p.~59]{Desoer1975_Book}, the part of the output corresponding to $u_{i,2}$ is also bounded, $\forall i=1,2$. From the standard superposition principle for linear systems, we have that $q_i-q_{c,i}\in\mathcal L_\infty$ and $\int_0^t [q_i(\sigma)-q_{c,i}(\sigma)]d\sigma\in\mathcal L_\infty$, $\forall i=1,2$. Based on the differentiation of (\ref{eq:27}) and (\ref{eq:28}), we have that $\dot \zeta_i^\ast\in\mathcal L_\infty$, $\forall i=1,2$. Using Property 1, we have from the third and fourth subsystems of (\ref{eq:41}) that $\dot s_i\in\mathcal L_\infty$, implying that $\ddot q_i\in\mathcal L_\infty$ and $\ddot \psi_i^\ast\in\mathcal L_\infty$, $\forall i=1,2$. Hence, $\dot \psi_i^\ast$ and $s_i$ are uniformly continuous, $\forall i=1,2$. From the properties of square-integrable and uniformly continuous functions \cite[p.~232]{Desoer1975_Book}, we have that $\dot \psi_i^\ast\to 0$ and $s_i\to0$ as $t\to\infty$, $\forall i=1,2$. From (\ref{eq:54}), we have that $\dot q_i-\dot q_{c,i}\in\mathcal L_\infty$, and hence $q_i-q_{c,i}$ is uniformly continuous, $\forall i=1,2$. From (\ref{eq:33}), (\ref{eq:34}), (\ref{eq:35}), and (\ref{eq:36}), we have that $\dot{\hat\vartheta}_i\in\mathcal L_\infty$ and $\dot{\hat w}_i\in\mathcal L_\infty$, which implies that $\hat\vartheta_i$ and $\hat w_i$ are uniformly continuous, $\forall i=1,2$. The result that $q_i-q_{c,i}\in\mathcal L_\infty$, $\dot q_i\in\mathcal L_\infty$,  $\ddot q_i\in\mathcal L_\infty$, $\dot\zeta_i^\ast\in\mathcal L_\infty$, $\psi_i^\ast\in\mathcal L_\infty$, $\dot \psi_i^\ast\in\mathcal L_\infty$, $\dot{\hat w}_{P,i}\in\mathcal L_\infty$, and $\dot{\hat w}_{I,i}\in\mathcal L_\infty$ implies that $\int_0^t[q_i(\sigma)-q_{c,i}(\sigma)]d\sigma$, $q_i$, $\dot q_i$, $\int_0^t\psi_i^\ast(\sigma)d\sigma$, $\psi_i^\ast$, $\zeta_i^\ast$, $\hat w_{P,i}$, and $\hat w_{I,i}$ are uniformly continuous, $\forall i=1,2$. It can be shown from (\ref{eq:37}), (\ref{eq:38}), (\ref{eq:39}), and (\ref{eq:40}) that $\dot{\hat w}_{P,i}$ and $\dot{\hat w}_{I,i}$ are uniformly continuous, $\forall i=1,2$. Using (\ref{eq:21}) and (\ref{eq:22}), we have that $\ddot z_i$ is piecewise uniformly continuous, $\forall i=1,2$. Hence, we have from the differentiation of (\ref{eq:27}) and (\ref{eq:28}) that $\dot \zeta_i^\ast$ is piecewise uniformly continuous, $\forall i=1,2$. It can then be demonstrated using Property 1 and from the third and fourth subsystems of (\ref{eq:41}) that $\dot s_i$ is piecewise uniformly continuous, $\forall i=1,2$. Therefore, $\ddot q_i$ and $\ddot \psi_i^\ast$ are piecewise uniformly continuous, $\forall i=1,2$. The application of the standard generalized Barbalat's lemma yields the conclusion that $\ddot \psi_i^\ast\to 0$, $\forall i=1,2$. For the system (\ref{eq:49}), using the input-output properties of uniformly marginally stable linear systems of the first kind \cite{Wang2020_AUT}, we have that $\dot \xi\to0$ as $t\to\infty$, and hence $\mathcal F_D(\xi)\to 0$ as $t\to\infty$. From the input-output properties of linear systems \cite[p.~59]{Desoer1975_Book}, we have by exploiting the relation $\ddot q_i=-\alpha \dot q_i+\dot \xi_i$, $\forall i=1,2$ [derived from (\ref{eq:23}) and (\ref{eq:24})] that $\dot q_i\to 0$ as $t\to\infty$, $\forall i=1,2$.
This then implies that $q_1-q_2(t-T_2)\to 0$ and $q_2-q_1(t-T_1)\to 0$ as $t\to\infty$. Exploiting the standard result that $q_i-q_i(t-T_i)=\int_0^{T_i}\dot q_i(t-\sigma)d\sigma\to 0$ as $t\to\infty$, $\forall i=1,2$, we have that $q_1-q_2(t-T_2)\to q_1-q_2\to 0$ and $q_2-q_1(t-T_1)\to q_2-q_1\to 0$ as $t\to\infty$.

Let us now analyze the torque reflection property of the teleoperator following the typical practice. In particular, consider the case that $\dot q_i$ and $\ddot q_i$ converge to zero, $i=1,2$. Based on (\ref{eq:1}) and (\ref{eq:2}), we have that $\tau_1^\ast\to K_{I,1}\int_0^t [q_1(\sigma)-q_{c,1}(\sigma)]d\sigma$ and $\tau_2^\ast\to -K_{I,2}\int_0^t [q_2(\sigma)-q_{c,2}(\sigma)]d\sigma$ and that $\dot q_{c,i}\to 0$ and $q_i-q_{c,i}\to 0$, $\forall i=1,2$. The first two subsystems of (\ref{eq:41}) can be rewritten as
\begin{align}
\label{eq:55}
\ddot \psi_1^\ast=&-\Lambda_{D,1}\dot\psi_1^\ast-\Lambda_{P,1}\psi_1^\ast-\Lambda_{I,1}\int_0^t \psi_1^\ast(\sigma)d\sigma\nn\\
&+\dot\xi_1+\lambda[\xi_1-\xi_2(t-T_2)]\\
\label{eq:56}
\ddot \psi_2^\ast=&-\Lambda_{D,2}\dot\psi_2^\ast-\Lambda_{P,2}\psi_2^\ast-\Lambda_{I,2}\int_0^t \psi_2^\ast(\sigma)d\sigma\nn\\
&+\dot\xi_2+\lambda[\xi_2-\xi_1(t-T_1)].
\end{align}
Using the result that $\dot q_i\to 0$ and $\ddot q_i\to 0$ and the exponential stability of (\ref{eq:a3}), we have from (\ref{eq:55}) and (\ref{eq:56}) that $\ddot \psi_i^\ast\to 0$, $\dot\psi_i^\ast\to 0$, and $\psi_i^\ast\to0$, $i=1,2$, and that $\int_0^t\psi_1^\ast(\sigma)d\sigma\to\lambda\alpha\Lambda_{I,1}^{-1}(q_1-q_2)$ and $\int_0^t\psi_2^\ast(\sigma)d\sigma\to\lambda\alpha\Lambda_{I,2}^{-1}(q_2-q_1)$. From equations (\ref{eq:31}) and (\ref{eq:32}), we have that
\begin{align}
\label{eq:57}
\dot q_{c,1}-\dot q_1\to&-\hat{\mathcal K}_{P,1} (q_{c,1}-q_1)\nn\\
&-\hat{\mathcal K}_{I,1}\int_0^t [q_{c,1}(\sigma)-z_1(\sigma)]d\sigma\\
\label{eq:58}
\dot q_{c,2}-\dot q_2\to&-\hat{\mathcal K}_{P,2} (q_{c,2}-q_2)\nn\\
&-\hat{\mathcal K}_{I,2}\int_0^t [q_{c,2}(\sigma)-z_2(\sigma)]d\sigma
\end{align}
which yields the result that $\hat{\mathcal K}_{I,i}\int_0^t[q_{c,i}(\sigma)-z_i(\sigma)]d\sigma\to 0$, $\forall i=1,2$. If $\hat w_{I,i}^{(k)}\ne 0$, we immediately have that $\int_0^t [q_{c,i}^{(k)}(\sigma)-z_i^{(k)}(\sigma)]d\sigma\to 0$, $\forall i=1,2,k=1,\dots,m$. If $\hat w_{I,i}^{(k)}\to 0$, then $\dot{\hat w}_{I,i}^{(k)} \to 0$; from (\ref{eq:39}) and (\ref{eq:40}), we also have that $\int_0^t[q_{c,i}^{(k)}(\sigma)-z_i^{(k)}(\sigma)]d\sigma\to0$ if $q_1^{(k)}-q_2^{(k)}\ne 0$ (implying that $s_i^{(k)}\ne 0$)\footnote{If $q_1^{(k)}-q_2^{(k)}=0$, it may be possible that the teleoperator stays at $\int_0^t \psi_i^{\ast(k)}(\sigma)d\sigma=0, \hat{w}_{I,i}^{(k)}=0,\dot{\hat w}_{I,i}^{(k)}=0,\int_0^t [q_i^{(k)}(\sigma)-q_{c,i}^{(k)}(\sigma)]d\sigma\ne 0$, yet the adaptation laws (\ref{eq:39}) and (\ref{eq:40}) involving $[\int_0^t \psi_i^{\ast(k)}(\sigma)d\sigma]^2$ renders this stay practically impossible.}, $\forall i=1,2,k=1,\dots,m$. Hence, $\int_0^t[q_{c,i}(\sigma)-z_i(\sigma)]d\sigma\to 0$, which implies that $\int_0^t[q_i(\sigma)-q_{c,i}(\sigma)]d\sigma \to \int_0^t [q_i(\sigma)-z_i(\sigma)]d\sigma$, $\forall i=1,2$.  It can then be shown that $\tau_1^\ast$ converges to $ K_{I,1}\Lambda_{I,1}^{-1}\lambda\alpha (q_1-q_2)$ and $\tau_2^\ast$ converges to $-K_{I,2}\Lambda_{I,2}^{-1}\lambda\alpha (q_2-q_1)$ stably provided that the linear system (\ref{eq:a3}) is exponentially stable, $\forall i=1,2$. This implies that $\tau_1^\ast\to K_{I,1}\Lambda_{I,1}^{-1}[K_{I,2}\Lambda_{I,2}^{-1}]^{-1}\tau_2^\ast$, i.e., the static torque quasi-reflection is guaranteed. The infinite manipulability of the teleoperator can be demonstrated using similar procedures as in the proof of Theorem 1. \hfill {\small $\blacksquare$}

\emph{Remark 3:} In the proof of Theorem 2, we do not require the assumption of uniform exponential stability concerning a linear time-varying system involving $\hat w_{P,i}$ and $\hat w_{I,i}$ via exploiting the input-output properties of an inverted form of the dynamics of robot manipulators with an inner PID control loop [see (\ref{eq:54})], in contrast with \cite[Theorem~4]{Wang2020_TAC}. In addition, a new Lyapunov-like function candidate that differs from the one in \cite{Wang2020_TAC} is empoloyed. This difference is due to the fact that the control objective for teleoperators is to achieve desired marginally stable dynamics (of the first kind) while the control objective for a single robot is to guarantee convergence of the regulation or tracking errors. The particular objective associated with teleoperators also motivates the controller design and stability analysis that are quite different from that for tracking or regulation of a single robot in \cite{Wang2020_TAC}; for instance, dynamic feedback exploiting the square-integrable functions that hold the possibility of being integral unbounded and input-output analysis for uniformly marginally stable linear time-varying systems constitute the essential part of the controller design and stability analysis.

\emph{Remark 4:} The condition given by (\ref{eq:42}) and (\ref{eq:43}) needs to be fulfilled so that the (marginal) stability of the teleoperator system can be ensured. It may seem challenging if considering the fact that the gains of the inner control loop set by the manufacturers, namely $K_{D,i}$, $K_{P,i}$, and $K_{I,i}$ are typically unknown or uncertain, $i=1,2$. On the one hand, in the theoretical sense, the condition given by (\ref{eq:42}) and (\ref{eq:43}) holds so long as $\gamma_i^\ast$ and $\gamma_i$ are large enough. On the other hand, we actually have some a priori knowledge of these gains by following some typical experience in designing PID controllers for second-order servoing systems, either from the perspective of stability or that of performance. In particular, consider a simple point mass under the action of a standard PID position controller
\be
m^\ast \ddot x=-k_D \dot x-k_P x-k_I \int_0^t x(\sigma)d\sigma
\ee
where $x\in R$ is the position of the point mass, $m^\ast\in R$ is the mass, and $k_D$, $k_P$, and $k_I$ are the derivative, proportional, and integral gains, respectively. From the standard linear system theory, we know that for ensuring the stability of the system, the gains are required to satisfy the condition that $k_Dk_P>m^\ast k_I$, leading to the standard engineering practice that $k_I$ is conservatively chosen to be very small while $k_P$ and $k_D$ are generally chosen to be much larger than $k_I$ and in addition, $k_P$ is chosen to be larger than $k_D$. Concerning the condition associated with robots with closed architecture given by (\ref{eq:42}) and (\ref{eq:43}), we would then have that $\lambda_{\max}\{K_{D,i}^{-1}K_{I,i}\}\ll 1$ and that $\lambda_{\max}\{K_{D,i}K_{P,i}^{-1}\}<1$ with $\lambda_{\max}\{\cdot\}$ denoting the maximum eigenvalue of a matrix, upon which the choice of $\gamma_i^\ast$ and $\gamma_i$ becomes quite convenient; for instance, one choice (might be conservative) is that $\gamma_i^\ast=1$ and $\gamma_i=\gamma_i^\ast$, $i=1,2$. In the reduced case of an inner PD controller, the condition given by (\ref{eq:42}) and (\ref{eq:43}) is no longer required; with $\gamma_i^\ast=0$, the design constant $\gamma_i$ can be chosen to be an arbitrary positive constant, $i=1,2$.

\emph{Remark 5:} A prominent point concerning the controller design for teleoperators with closed architecture is the increment of complexity due to the PID control action in the inner loop in comparison with the reduced case of an inner PD controller (in which case, the control law can be straightforwardly derived following similar procedures as in the case of an inner PID controller); in addition, the condition for ensuring the stability of the closed-loop system also becomes much involved due to the uncertainty of $\mathcal K_{P,i}$ and $\mathcal K_{I,i}$, $i=1,2$. While an inner PID position controller provides many advantages in the case of position regulation or tracking of a single robot, the study here, on the other hand, shows the challenge and difficulties in designing bilateral controllers for teleoperator systems with an inner PID position controller. The robotics industry is increasingly expanding and the suggestion we propose specifically to robot manufacturers is that the OPTION that the integral action of the inner control loop can be removed is provided to the user or that the ratios of the damping, proportional, and integral gains (namely $\mathcal K_{P,i}$ and $\mathcal K_{I,i}$, $i=1,2$) can be accessed so as to facilitate the expansion of application scenarios of industrial/commercial robots (for instance, human-robot interaction).

\section{Dynamic Separation and a New Class of Kinematic Control}

Separation property is an important guideline for designing kinematic controllers for robots with closed architecture (see, e.g., \cite{Wang2017_TAC}), which is typically formulated in the context of standard cascaded systems. The dynamics (\ref{eq:41}) yielded by the adaptive dynamic controller in Sec. IV exhibit the dynamic-cascaded property, and to accommodate this issue, we formulate a new separation property which is referred to as dynamic separation. The dynamic separation here extends the separation (or more precisely ``static separation'') in \cite{Wang2017_TAC}. Specifically, even if the first two subsystems of (\ref{eq:41}) involve the derivative of the state of the third and fourth subsystems of (\ref{eq:41}), the separation is still achieved. This dynamic separation yields a kinematic controller for teleoperators with an inner PID position controller.

\emph{Theorem 3:} Suppose that the inner PID control loop can guarantee that the joint velocity and position tracking errors and the integral of the joint position tracking error are square-integrable and bounded. The kinematic controller given by $\dot q_{c,i}^\ast=\dot z_i$, $i=1,2$ given by (\ref{eq:21}) and (\ref{eq:22}) with $z_1=q_{c,1}$ and $z_2=q_{c,2}$ for the teleoperator system given by (\ref{eq:1}) and (\ref{eq:2}) with the gravitational torques being exactly compensated a priori ensures the position synchronization of the master and slave robots in free motion, and static torque quasi-reflection under the condition that the linear system (\ref{eq:a3}), $\forall i=1,2$, is exponentially stable. In addition, the teleoperator system is infinitely manipulable with degree one.

The proof of Theorem 3 can be directly performed by following similar procedures as in that of Theorem 2. The benefit of the kinematic controller yielded by the dynamic separation property is that design flexibility is enlarged as compared with the kinematic controller yielded by the static separation given by (\ref{eq:9}) and (\ref{eq:10}), and that it does not involve the discontinuity of the time-varying delay, which thus potentially improves the performance of the teleoperator system (for instance, enhanced robustness and smoother dynamic response even in the case of discontinuous delay).

\emph{Remark 6:} The kinematic control yielded by using dynamic separation is favorable in the sense that the design freedom is enlarged and that the joint velocity command is differentiable even if the time-varying delays are discontinuous. In addition, the kinematic control in the sense of dynamic separation potentially ensures better performance (e.g., better torque reflection) due to the fact that the order of such control is increased; see the kinematic controller given by (\ref{eq:21}) and (\ref{eq:22}) as compared with the one given by (\ref{eq:9}) and (\ref{eq:10}).

\section{Experimental Results}

In this section, we consider the application of the proposed adaptive control to a teleoperator system involving a Phantom Omni and a UR10 robot (see Fig. 1 and Fig. 2), and only the first three DOFs (degrees of freedom) of the Phantom Omni and UR10 are controlled where the linear transformations concerning the joint position measurements are performed so that the joint position synchronization of the Phantom Omni and UR10 is convenient for direct manipulation. The control architecture of the teleoperator system is hybrid in the sense that the Phantom Omni has an open torque design interface while the torque design interface of the UR10 robot is unavailable (an inner PID position controller is expected to be embedded, as is the typical case in most industrial/commercial robots). The choice of such hybrid teleoperator systems for performing the experimental study is believed to be representative since in most scenarios, the master robot under the action of the human operator is typically a joystick (for instance, Phantom Omni) and the slave robot is an industrial/commercial robot that can conduct various operations required by specific engineering applications. The teleoperator system involves contact with a table (as the environment), as shown in Fig. 2. The forward and backward time delays $T_1(t)$ and $T_2(t)$ are set to conform to uniform distributions over the interval $[0.3,0.9]$ (s) with $T_1$ being updated every 96 ms and $T_2$ every 100 ms.

\begin{figure}
\centering
%%----start of first figure----
\begin{minipage}[t]{1.0\linewidth}
\centering
\includegraphics[width=2.6in]{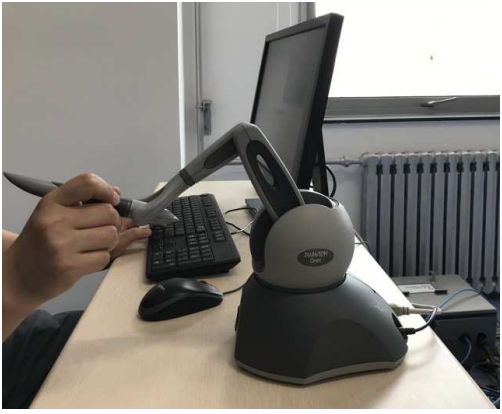}
\caption{The Phantom Omni.}
\end{minipage}%
\end{figure}

\begin{figure}
\centering
%%----start of first figure----
\begin{minipage}[t]{1.0\linewidth}
\centering
\includegraphics[width=2.6in]{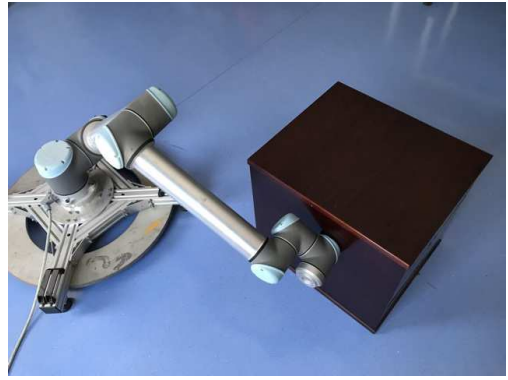}
\caption{The UR10 robot and environment (a table).}
\end{minipage}%
\end{figure}

The Phantom Omni (as the master) has an open torque design interface with the controller in \cite{Wang2020b_AUT} being used, namely
\begin{align}
\begin{cases}
\dot z_1=-\alpha \dot q_1-\lambda[\xi_1-\xi_2(t-T_2)]+\lambda_{\mathcal M}(\dot q_1-z_1)\\
s_1=\dot q_1-z_1\\
\tau_1=-K_1 s_1+Y_1(q_1,\dot q_1,z_1,\dot z_1)\hat\vartheta_1\\
\dot{\hat \vartheta}_1=-\Gamma_1 Y_1^T (q_1,\dot q_1,z_1,\dot z_1) s_1
.\end{cases}
\end{align}
The controller parameters for the Phantom Omni are chosen as $\alpha=1.5$, $\lambda_{\mathcal M}=2$, $\lambda=36$, $K_1=0.02 I_3$, and $\Gamma_1=0.0005 I_{12}$. The initial value of $\hat\vartheta_1$ is set as $\hat\vartheta_1(0)=0_{12}$, and the initial value of $z_1$ is set as $z_1(0)=[0,0,0]^T$. The gravitational torque is a priori compensated rather than adaptively compensated, and this is for ensuring the infinite manipulability of the teleoperator system. The sampling period for the Phantom Omni is set as 1 ms.

The UR10 robot (as the slave), on the other hand, has closed architecture and the controller (with the joint velocity command as the control input) is given by (\ref{eq:32}), (\ref{eq:34}), (\ref{eq:36}), (\ref{eq:38}), and (\ref{eq:40}) with $z_2$, $\dot z_2$, and $\ddot z_2$ being given by (\ref{eq:22}). The controller parameters are chosen as $\alpha=1.5$, $\Lambda_{D,2}=15 I_3$, $\Lambda_{P,2}=75 I_3$, $\Lambda_{I,2}=125 I_3$, $\lambda=20$, $\gamma_2=30$, $\gamma_2^\ast=30$, $\Gamma_2=0.3 I_{12}$, $\Gamma_2^\ast=0.005 I_3$, $\Gamma_{P,2}^\ast=10 I_3$, and $\Gamma_{I,2}^\ast=10 I_3$. It is expected by the standard engineering practice that the choice of $\gamma_2$ and $\gamma_2^\ast$ here satisfies the condition given by (\ref{eq:42}) and (\ref{eq:43}). The initial values of $\hat\vartheta_2$, $\hat w_2$, $\hat w_{P,2}$, and $\hat w_{I,2}$ are, respectively, set as $\hat\vartheta_2(0)=0_{12}$, $\hat w_2(0)=[0,0,0]^T$, $\hat w_{P,2}(0)=[3,3,3]^T$, and $\hat w_{I,2}(0)=[0.5,0.5,0.5]^T$. The initial values of $z_2$ and $\dot z_2$ are set as $z_2(0)=q_2(0)$ and $\dot z_2(0)=[0,0,0]^T$, respectively. The gravitational torque is not compensated by the adaptive controller since the inner control loop of UR10 has performed the gravitational torque compensation. The sampling period for updating the joint velocity command is set as 8 ms.

The experimental results are shown in Fig. 3 to Fig. 8. This shows that the positions of the Phantom Omni and UR10 synchronize in free motion even if the communication channel is subjected to delays that vary quite fast and involve discontinuous points (as is demonstrated in Fig. 3, Fig. 4, and Fig. 5); in the contact scenario, the proposed control exhibits the static torque-quasi-reflecting property, as shown in Fig. 6, Fig. 7, and Fig. 8 where $-\tau_1$ and $-{\rm diag}[3,3.25,3.75]\int_0^t [q(\sigma)-q_c(\sigma)]d\sigma$ are given and respectively converge to $\tau_1^\ast$ and a scale of $-\tau_2^\ast$ as the teleoperator approaches the static state (this practice is preferred in the case without direct interaction torque measurement \cite{Lee2010_TRO_PassiveSetPosition}).

\begin{figure}
\centering
%%----start of first figure----
\begin{minipage}[t]{1.0\linewidth}
\centering
\includegraphics[width=2.6in]{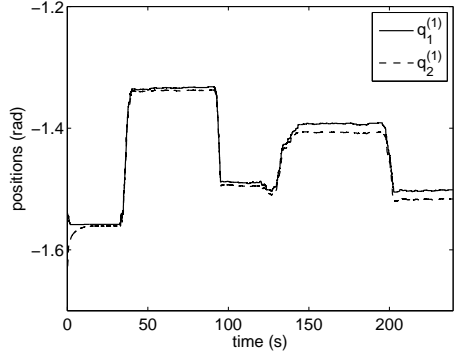}
\caption{Positions of the Phantom Omni and UR10 (first coordinate).}
\end{minipage}%
\end{figure}

\begin{figure}
\centering
%%----start of first figure----
\begin{minipage}[t]{1.0\linewidth}
\centering
\includegraphics[width=2.6in]{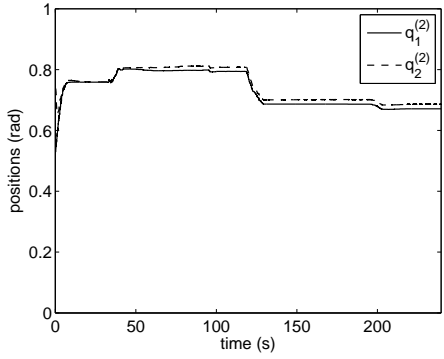}
\caption{Positions of the Phantom Omni and UR10 (second coordinate).}
\end{minipage}%
\end{figure}

\begin{figure}
\centering
%%----start of first figure----
\begin{minipage}[t]{1.0\linewidth}
\centering
\includegraphics[width=2.6in]{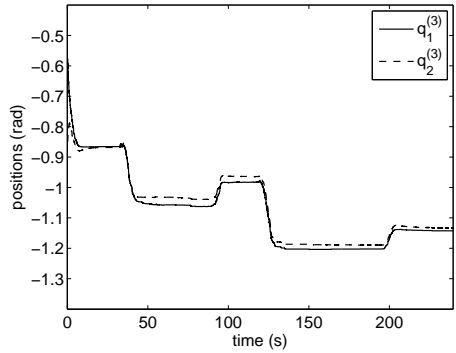}
\caption{Positions of the Phantom Omni and UR10 (third coordinate).}
\end{minipage}%
\end{figure}

\begin{figure}
\centering
%%----start of first figure----
\begin{minipage}[t]{1.0\linewidth}
\centering
\includegraphics[width=2.6in]{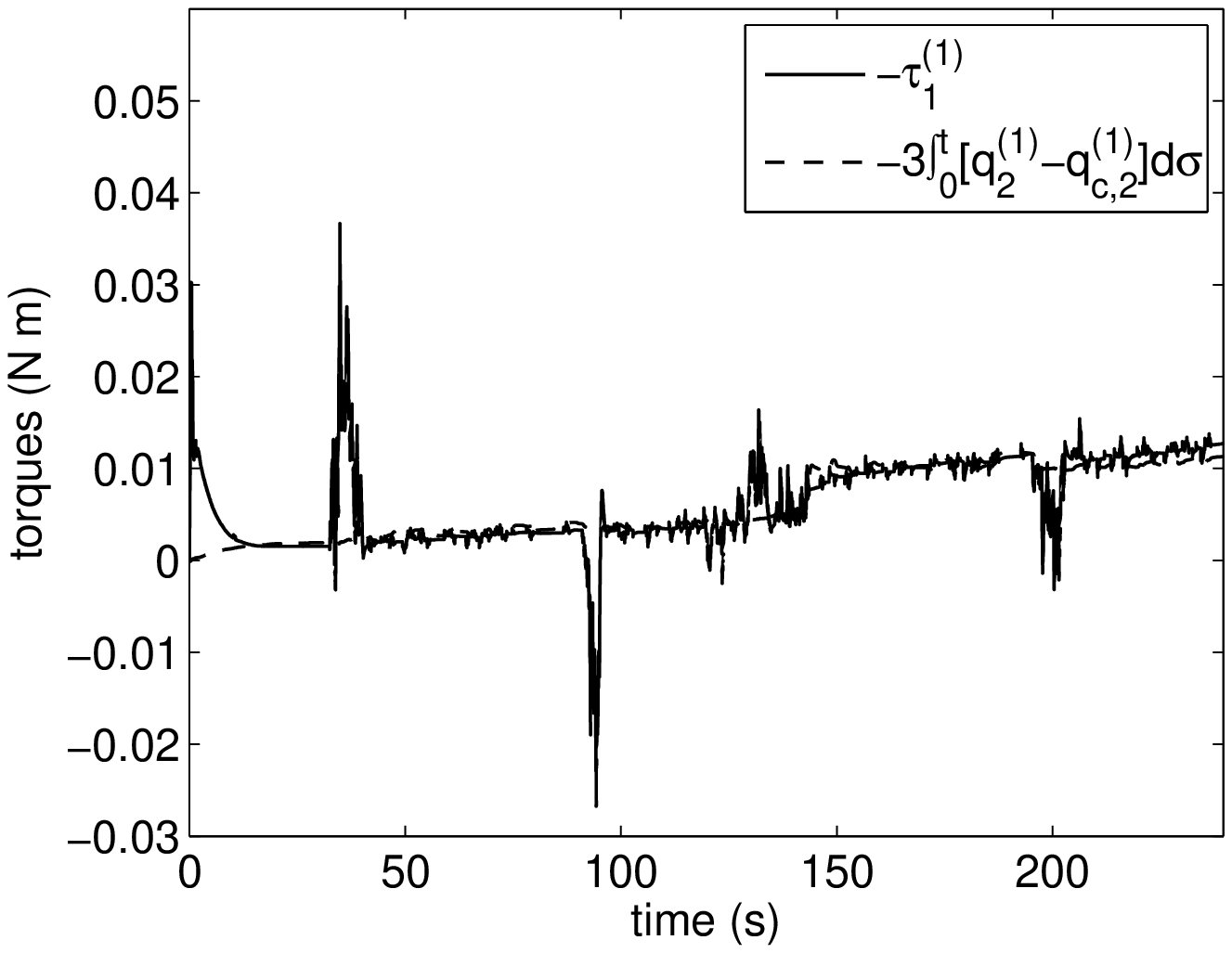}
\caption{The response of $-\tau_1^{(1)}$ and $-3\int_0^t[q_2^{(1)}(\sigma)-q_{c,2}^{(1)}(\sigma)]d\sigma$.}
\end{minipage}%
\end{figure}

\begin{figure}
\centering
%%----start of first figure----
\begin{minipage}[t]{1.0\linewidth}
\centering
\includegraphics[width=2.6in]{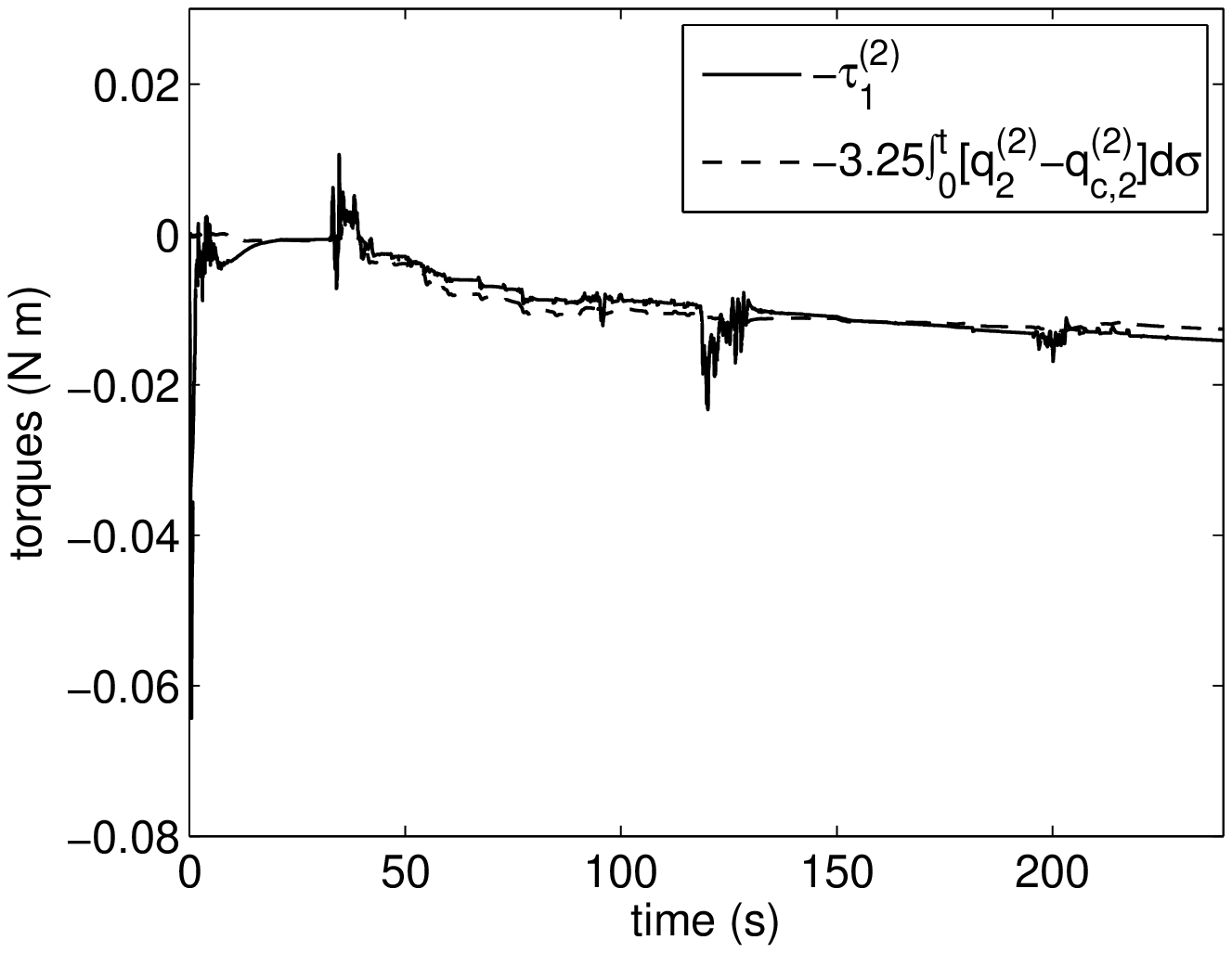}
\caption{The response of $-\tau_1^{(2)}$ and $-3.25\int_0^t[q_2^{(2)}(\sigma)-q_{c,2}^{(2)}(\sigma)]d\sigma$.}
\end{minipage}%
\end{figure}

\begin{figure}
\centering
%%----start of first figure----
\begin{minipage}[t]{1.0\linewidth}
\centering
\includegraphics[width=2.6in]{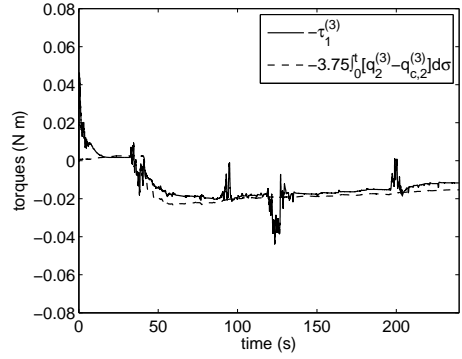}
\caption{The response of $-\tau_1^{(3)}$ and $-3.75\int_0^t[q_2^{(3)}(\sigma)-q_{c,2}^{(3)}(\sigma)]d\sigma$.}
\end{minipage}%
\end{figure}

\section{Conclusion}

This paper provides delay-independent solutions to the bilateral control problem of teleoperators involving robots with closed architecture and subjected to arbitrary bounded time-varying delay. We propose kinematic and adaptive dynamic controllers based on a new class of dynamic feedback, and in addition, the proposed adaptive dynamic control takes into account the uncertainty and dynamic effects of the inner control loop. It is shown that the proposed controllers ensure both the infinite manipulability of the teleoperator system with degree one and robustness with respect to arbitrary bounded unknown time-varying delay, and the position synchronization in free motion and static torque quasi-reflection are also ensured by the proposed controllers. %The simulation results and experimental results using a teleoperator system consisting of a Phantom Omni (with an open torque design interface) and a UR10 robot (with closed architecture) are provided to show the performance of the proposed controllers based on dynamic feedback.

It is well known that robots with closed architecture (e.g., most industrial/commercial robots) typically employ an inner PID position control action. With respect to the velocity tracking error, the proportional action and integral action are equivalent to the integral action and double-integral action, respectively. While this is favorable for stabilization of the tracking or regulation errors, it brings challenges for bilateral teleoperation since adding the integral action generally weakens the manipulability of a teleoperator system. By exploiting the properties of a class of square-integrable functions that are probable to be integral unbounded, we develop kinematic and dynamic controllers which can guarantee that the teleoperator systems with closed architecture are infinitely manipulable with degree one.

% or


%\appendices

%\section{}

\section*{Acknowledgment}

The authors would like to thank Dr. Yang Zhou for the
valuable discussions on this topic and for the help concerning the experimental study.

%\ifCLASSOPTIONcaptionsoff
%  \newpage
%\fi



\bibliographystyle{IEEEtran}
\bibliography{..//Reference_list_Wang}
%\begin{thebibliography}{1}
%
%
%\end{thebibliography}

%\begin{biography}[{\includegraphics[width=1in,height=1.25in,clip,keepaspectratio]{mshell}}]{Michael Shell}

%\end{IEEEbiography}


%\newpage



%\vfill


\end{document}